%% file: paper.tex
\documentclass[aps,prl,twocolumn,showpacs,superscriptaddress,preprintnumbers]{revtex4}  
\usepackage{graphicx}  
\usepackage{dcolumn}   
\usepackage{bm}        
\usepackage{amssymb}   
\usepackage{xspace}
\usepackage{multirow}
\usepackage{color}
\usepackage{units}
\usepackage{endnotes}

\newcommand{\etal}{{\it et al.}}

\newcommand{\RHCpot}{\ensuremath{\unit[3.36\times10^{20}]{POT}}\xspace}

\newcommand{\FHCnounit}{\mbox{\ensuremath{10.71 \times 10^{20}\xspace}}}

\newcommand{\delmsq}[1]{\ensuremath{\Delta m^2_{ #1 }}\xspace}
\newcommand{\dmsq}{\ensuremath{\Delta m^2}\xspace}
\newcommand{\dmbarsq}{\ensuremath{\Delta \overline{m}^2}\xspace}

\newcommand{\sinsq}{\ensuremath{\sin^{2}\!\left(2\theta \right)}\xspace}
\newcommand{\sinsqbar}{\ensuremath{\sin^{2}\!\left(2\overline{\theta} \right)}\xspace}

\newcommand{\numu}{\ensuremath{\nu_{\mu}}\xspace}          
\newcommand{\numubar}{\ensuremath{\overline{\nu}_{\mu}}\xspace}   

\newcommand{\nue}{\ensuremath{\nu_{e}}\xspace}                      
\newcommand{\nuebar}{\ensuremath{\overline{\nu}_{e}}\xspace}        

\newcommand {\nubar}{\ensuremath{\overline{\nu}}\xspace}

\newcommand{\sntwoparallbarenum}{\ensuremath{0.950}}
\newcommand{\sntwoparall}{\ensuremath{\sntwoparallbarenum^{+0.035}_{-0.036}}\xspace}
\newcommand{\sntwoparalllimit}{\ensuremath{0.890}\xspace}

\newcommand{\dmsqtwoparallbarenum}{\ensuremath{2.41}}
\newcommand{\dmsqtwoparallnounit}{\ensuremath{\dmsqtwoparallbarenum^{+0.09}_{-0.10}}}
\newcommand{\dmsqtwoparall}{\ensuremath{\unit[(\dmsqtwoparallnounit) \times 10^{-3}]{eV^{2}}}\xspace}

\newcommand{\snbarfourparall}{\ensuremath{0.97^{+0.03}_{-0.08}}\xspace}
\newcommand{\snbarfourparalllimit}{\ensuremath{0.83}\xspace}
\newcommand{\dmbarsqfourparall}{\ensuremath{\unit[(2.50 ^{+0.23}_{-0.25}) \times 10^{-3}]{eV^{2}}}\xspace}
\newcommand{\dmbarsqdifffourparall}{\ensuremath{\unit[(0.12^{+0.24}_{-0.26}) \times 10^{-3}]{eV^{2}}}\xspace}

\newcounter{rowcounter}
\addtocounter{rowcounter}{-1} 

\begin{document}


\title{Measurement of Neutrino and Antineutrino Oscillations Using Beam and Atmospheric Data in MINOS}

\input{NuMI-Atm-NuOsc-Sep12-authors.tex} 
\date{\today}
\preprint{FERMILAB-PUB-13-102-PPD}
\preprint{arXiv:hep-ex/1304.6335}

\begin{abstract}
We report measurements of oscillation parameters from $\numu$
and $\numubar$ disappearance using beam and atmospheric data from
MINOS. The data comprise exposures of \unit[\FHCnounit]{protons on
  target (POT)} in the $\nu_{\mu}$-dominated beam, \RHCpot in the $\numubar$-enhanced beam, and 37.88 kton-years of atmospheric neutrinos.  Assuming identical $\nu$ and $\nubar$ oscillation parameters, we measure \mbox{$|\dmsq| = \dmsqtwoparall$} and $\sinsq = \sntwoparall$. Allowing independent $\nu$ and $\nubar$ oscillations, we measure antineutrino parameters of $|\dmbarsq| = \dmbarsqfourparall$ and
$\sinsqbar = \snbarfourparall$, with minimal change to the neutrino parameters.
\end{abstract}

\maketitle

Neutrino oscillation provides direct evidence that neutrinos have
non-zero mass and represents the only phenomenon observed to date with
an origin beyond the Standard Model of particle
interactions. 
With massive neutrinos, three flavor eigenstates mix with three mass eigenstates according to a unitary matrix that can be parameterized by three angles and a CP-violating phase~\cite{ref:PDG2012}.
The resulting oscillation probability depends on the mixing angles and
on the differences between the squared neutrino masses. 
The MINOS experiment performs precision measurements of oscillations via $\numu$
disappearance.  These oscillations are well described by an effective
two-flavor model with flavor and mass eigenstates related by a single
mixing angle $\theta$.  In this approximation, the $\numu$ survival
probability is given by 
\[
P \left( \nu_{\mu} \rightarrow \nu_{\mu} \right) = 1-\mbox{sin}^{2} 2\theta \mbox{ } \mbox{sin}^{2} \left( \frac{1.267 \mbox{ } \Delta m^{2} [\mbox{eV}^{2}] \mbox{ } L [\mbox{km}] }{ E[\mbox{GeV}] } \right),
\] 
where $L$ is the distance traveled by the neutrino and $E$ is its
energy.  The mass splitting, \dmsq, is an
admixture of  the three-flavor parameters $\delmsq{31}$ and $\delmsq{32}$~\cite{ref:dm2note}, and
it 
differs from $\delmsq{32}$ by less than 2\%. The \numubar survival probability has the same form, but the mixing parameters are denoted by \dmbarsq and \sinsqbar.

  
The MINOS measurements use neutrinos produced in the NuMI accelerator complex and by cosmic ray interactions in the atmosphere. The accelerator provides a source of
neutrinos with a fixed baseline and an energy spectrum that peaks at $L/E\sim\unit[250]{km/GeV}$, close to the region where the $\numu$ survival probability reaches its first minimum. Atmospheric neutrinos are produced
with a broad range of $\unit[E(\sim0.5-10^{4}]{GeV)}$ and $\unit[L(\sim10-10^{4}]{km)}$,
enabling the study of oscillations across a wide region
in $L/E$ and covering many oscillation cycles.
The precision of the oscillation measurement is enhanced by combining these two complementary samples.

This Letter presents the first ever joint analysis of atmospheric
  and accelerator neutrinos in the same experiment. The new results come from the full
  MINOS data set, collected over a period of nine years.
  The combination of data sets, together with increased exposures,
  produces a significant improvement in the sensitivity to
  oscillations over previous MINOS analyses~\cite{ref:minosCC2010,ref:minosAtmos2012,ref:minosRHC_2.95,ref:minosFHCNuBars}. Furthermore,
  MINOS has the unique ability to separate neutrinos and antineutrinos on an event-by-event basis. Coupled with the world's only set of long-baseline accelerator antineutrino data, we present the most precise measurements to date of the larger mass splitting for both neutrinos and antineutrinos.


The NuMI beam~\cite{ref:beam2} is produced at the Fermi National Accelerator Laboratory (Fermilab) by
\unit[120]{GeV} protons striking a graphite target. The resulting
charged pions and kaons are focused by two magnetic horns before decaying in a \unit[675]{m} long
helium-filled volume~\cite{ref:HeNote}. 
The beam is directed through a hadron absorber and rock to stop all
particles except neutrinos.
The energy spectrum of the neutrino beam can be changed by varying the distance between the target and first horn.
Most of the data
used in this analysis were collected with a spectrum peaking at a
neutrino energy of \unit[3]{GeV}. By selectively focusing positive or
negative pions and kaons, a  \numu-dominated or \numubar-enhanced beam
is produced.

The two MINOS detectors are steel-scintillator tracking calorimeters with toroidal magnetic fields~\cite{ref:nim}.  Each detector consists of steel plates with segmented plastic scintillator planes mounted on them. The planes are perpendicular to the beam direction.  The \unit[0.98]{kton} Near Detector (ND), located \unit[1.04]{km} from the neutrino production target, measures the beam composition and energy near the source.  
The \unit[5.4]{kton} Far Detector (FD) measures the beam composition and energy spectrum \unit[735]{km} away from the target.  Installed \unit[705]{m} (\unit[2070]{m} water-equivalent) underground in the Soudan Underground Laboratory in Minnesota, the FD is also used to measure oscillations in atmospheric neutrinos and antineutrinos. A scintillator veto shield is installed above the FD to enhance the rejection of the cosmic-ray muon background.

Muon neutrinos and antineutrinos are identified through their charged
current (CC) interactions \[\numu(\numubar)+X\rightarrow
\mu^{-}(\mu^{+})+X^{\prime}.\] 
The muon typically deposits energy in the detector in a clear track-like pattern.  The hadronic recoil system, $X^{\prime}$, leaves a diffuse shower-like deposition pattern.
The only notable background in the CC sample arises from a small
  number of neutral current (NC) interactions that generate only
  hadronic activity but can display a track-like signature.
Muon neutrinos and antineutrinos are separated by the direction of curvature of the charged muon track in the magnetic field of the detectors.
The muon momentum is determined from the range for muons that stop in the detector and from curvature for exiting muons. 
For beam neutrino interactions, a $k$-Nearest-Neighbor classification algorithm ($k$NN) is used to estimate the hadronic energy from both the calorimetric energy deposited and the topology of the shower~\cite{ref:ChrisThesis}.   For atmospheric neutrino interactions, the calorimetric energy deposits in each scintillator strip are summed to provide an estimate of the true shower energy. The reconstructed neutrino energy is given by the sum of the muon and shower energy measurements.

Our new results are based on FD exposures of
\unit[\FHCnounit]{protons on
  target (POT)} in the \numu-dominated beam and
 \RHCpot in the \numubar-enhanced beam,
corresponding to increases of 48\% and 14\%,
respectively, over our previous analyses~\cite{ref:minosCC2010, ref:minosRHC_2.95}.
As in these previous analyses, the selection of \numu and
\numubar CC interaction candidates proceeds via the
construction of a set of variables that characterize
the event topology and energy deposition of
muon tracks~\cite{ref:RustemThesis}. Again, these variables are combined
into a single discriminating variable using
a $k$NN technique.

From the \numu-dominated beam, we use both neutrinos and antineutrinos
with interaction vertices contained within the detectors' fiducial volumes. When
explicitly fitting antineutrino oscillation parameters, we apply an
optimized antineutrino event selection to increase the purity of this
contained-vertex antineutrino sample~\cite{ref:minosFHCNuBars}.  
In this beam mode, we also select a sample of non-fiducial muons in the FD, comprising \numu CC and \numubar CC interactions outside the fiducial volume and in the rock surrounding the detector~\cite{ref:MattStraitThesis,
  ref:AaronThesis}. For such interactions, the muon energy alone is
used as the neutrino energy estimator. No muon charge-sign separation is performed on this sample
since many of these muons are confined to the edges of the detector
where the magnetic field is very low and muon curvature is less well
modeled. In the \numubar-enhanced beam, only the contained-vertex
antineutrino sample is used, as the non-fiducial sample is dominated by high energy neutrinos. 

%
The predicted FD beam spectra are derived from the observed ND beam data using a beam transfer matrix~\cite{ref:minos2008}. This extrapolation procedure mitigates
many sources of systematic uncertainties and naturally accounts for any
variations in the beam conditions such as target degradation or
differences among the seven different production targets used throughout the experiment's lifetime.  Since the ND is used to provide a baseline
spectrum, it is important to minimize any differences
between the response in the two detectors.  In particular, the
region around the ND magnetic coil is poorly modeled, so any beam-induced events with
muon tracks entering this region are removed from the ND data set.

We use the same atmospheric neutrino data set
    and event samples as our previous analysis~\cite{ref:minosAtmos2012},
    which is based on a FD exposure of 37.88 kton-years.
    The events are identified by the presence of either an
    interaction vertex within the fiducial volume of the
    detector or an upward-going or horizontal muon track.
    The selected events are separated into three samples:
contained-vertex muons, non-fiducial muons, and contained-vertex showers.
The two muon samples are produced
by $\numu$ CC and $\numubar$ CC interactions; the contained-vertex shower
sample is
composed mainly of $\nue$ CC, $\nuebar$ CC and NC interactions.
The atmospheric neutrino samples must be selected from a background of
cosmic-ray muons. For contained-vertex muons and showers, this
background is reduced to 4\% and 12\%, respectively, by applying a series of containment requirements and by checking for energy deposits in the sections of veto shield above the fiducial event vertex within a \unit[$\pm50$]{ns} window. 
In the non-fiducial sample, the background is almost entirely removed by using the \unit[2.5]{ns} timing resolution to accurately determine the incoming muon track direction.
 Table~\ref{tab:evtsel} lists the numbers of observed events
  and the corresponding predictions, with and without oscillations,
  for each of the analyzed samples. 

\begin{figure*}[htb]
\centering
\includegraphics[width=\textwidth]{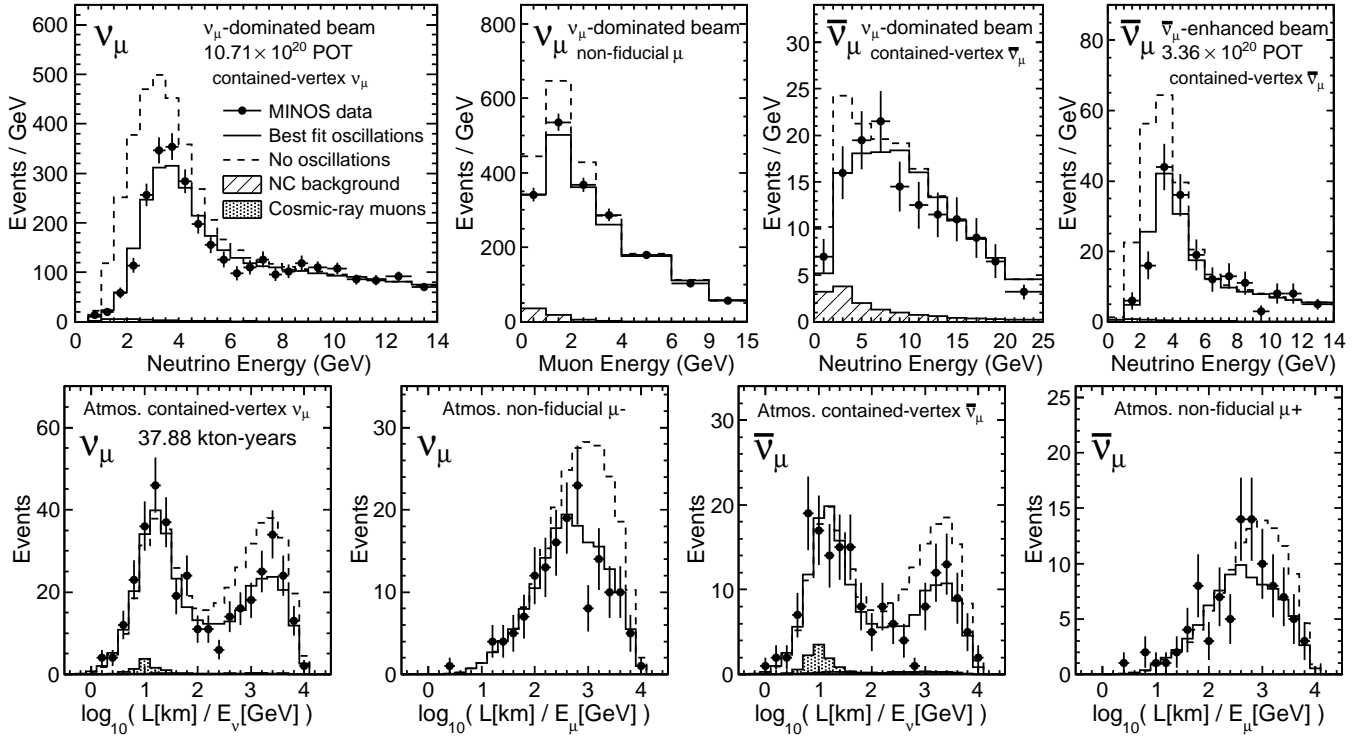}
\caption{FD data samples compared to predictions with and without oscillations.  The top row shows the energy spectra of the beam samples, while the bottom row shows the L/E distributions for the atmospheric event samples.
\label{fig:CCOnlyESpec}}
\end{figure*} 

We simulate atmospheric neutrino events according to the Bartol flux
calculations~\cite{ref:Bartol}.  The beam neutrino flux is simulated using the 
FLUGG~\cite{ref:FlukaAIP}
package, which combines 
GEANT4~\cite{ref:GEANT4IEEE} 
geometry with 
FLUKA~\cite{ref:FlukaAIP} 
hadron production.  All beam neutrino interactions, and interactions of atmospheric neutrinos in the detectors, are simulated using NEUGEN3~\cite{ref:Neugen}.  
We simulate atmospheric neutrino interactions in the surrounding rock using NUANCE~\cite{ref:Nuance}, which propagates the final-state muons through the rock and up to the edge of the detector.
In both the beam and atmospheric simulations, the propagation of
particles in the detector, and the detector response, are simulated
with
GCALOR~\cite{ref:GCALOR} and GEANT3~\cite{ref:GEANT3}. The simulation
incorporates the background arising from $\numu \rightarrow \nu_{\tau}$ appearance. For the best fit oscillations, the predicted event yield from this channel totals 18 events across the entire data set.

The oscillation parameters are obtained from a maximum likelihood fit
to the data. The measured FD beam data are binned as a function of reconstructed
neutrino energy.   
To improve the sensitivity of the analysis, the contained-vertex \numu
events from the \numu-dominated beam are 
divided into five sub-samples according to their estimated energy
resolution, which is calculated from their measured muon and shower
energies and lies primarily in the range $5-30\%$~\cite{ref:StephenThesis,ref:JessThesis,ref:minosCC2010}.
The atmospheric samples are binned as a function of $L/E$.
 The contained-vertex atmospheric \numu and \numubar events are divided
into four sub-samples according to the estimated $L[\mathrm{km}]/E[\mathrm{GeV}]$ resolution,
where $\sigma_{\log_{10}(L/E)}$ ranges from 0.05 to 1.2~\cite{ref:minosAtmos2012NIM}.
The contained-vertex showers are grouped in a single bin because the
majority are too short for an accurate measurement of $L/E$.  These events are relatively insensitive to
oscillations but provide a constraint on the overall normalization of the atmospheric flux.

\begin{table}
\begin{tabular}{l c c c} 
\hline
              &  \multicolumn{2}{c}{Simulation} & Events \\
Data Set & No osc. & With osc.  & Observed \\
\hline
\numu from \numu beam                          & 3201 & 2543 & 2579 \\
\numubar from \numu beam                     & 363   & 324   & 312 \\
Non-fiducial $\mu$ from \numu beam & 3197 & 2862 &  2911 \\
\numubar from \numubar beam                & 313   & 227   &  226  \\
Atm. contained-vertex \numu + \numubar& 1100& 881   & 905 \\
Atm. non-fiducial $\mu^-$ + $\mu^+$ & 570  & 467  & 466 \\
Atm. showers                                               & 727  & 724  & 701\\
\hline
\end{tabular}
\caption{Numbers of events selected in each sample. The oscillated
  event yields come from the best fit to all data, assuming identical
  $\nu$ and $\nubar$ oscillations (\mbox{$\unit[|\dmsq| = \dmsqtwoparallbarenum\times 10^{-3}]{eV^{2}}$} and $\sinsq = \sntwoparallbarenum$). \label{tab:evtsel}
}

\end{table}

  The fit incorporates a set of nuisance parameters
that accommodate the largest
systematic uncertainties 
in the simulation of the beam~\cite{ref:minosCC2010,ref:minosRHC_2.95} and the atmospheric~\cite{ref:minosAtmos2012} neutrino data.
For both data sets,
the fit incorporates the systematic uncertainties
in the overall normalizations of the event samples,
the relative normalization of the NC background component, the muon
momentum, and the shower energy.
The latter two uncertainties are taken as correlated
between the beam and atmospheric samples. An analysis
performed with all uncertainties uncorrelated produces
similar results. Additional systematic parameters are
included in the fit to cover the uncertainties in the
rate and spectral shape of atmospheric \numu and \numubar
events arising from uncertainties in the neutrino
flux and cross-section simulations. 





\begin{figure}[htb]
\centering
\includegraphics[width=\columnwidth]{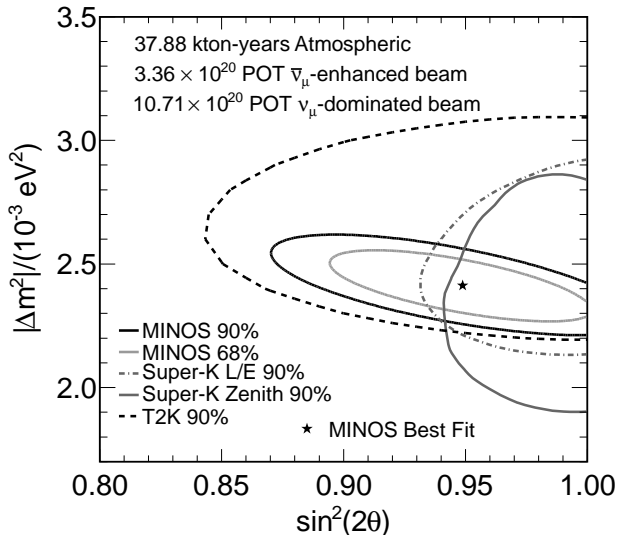}
\caption{The allowed regions of $|\dmsq|$ and
  \sinsq, assuming identical neutrino and antineutrino
  oscillations. The MINOS result is compared to results from
Super-Kamiokande~\cite{ref:SKNeutrino2012} and T2K~\cite{ref:T2Knucontour}. }
\label{fig:nu}
\end{figure}

When we fit the full MINOS data sample to the two-flavor neutrino oscillation hypothesis, assuming that neutrinos and
antineutrinos have identical oscillation parameters, we obtain $|\dmsq|
= \dmsqtwoparall$ and $\sinsq = \sntwoparall$. Maximal mixing is 
disfavored at the 86\% confidence level (C.L.); we measure $\sinsq  > \sntwoparalllimit$ at 90\% C.L. 
The observed beam and atmospheric event spectra in the
    FD are shown in Fig.~\ref{fig:CCOnlyESpec}, along with the predictions for
    the case of no oscillations and the best fit.
The data are well described by the neutrino oscillation model; the same analysis performed on simulated experiments returns a worse quality of fit for 19.1\% of those experiments.  A number of cross checks were performed by fitting each of the data samples separately.  Those separate fits yielded consistent oscillation parameters, indicating that the data samples are consistent with each other and with the oscillation hypothesis. 
Allowed regions for the oscillation parameters, assuming identical neutrino and antineutrino oscillations, are shown in Fig.~\ref{fig:nu}.  

\begin{figure}[htb]
\centering
\includegraphics[width=\columnwidth]{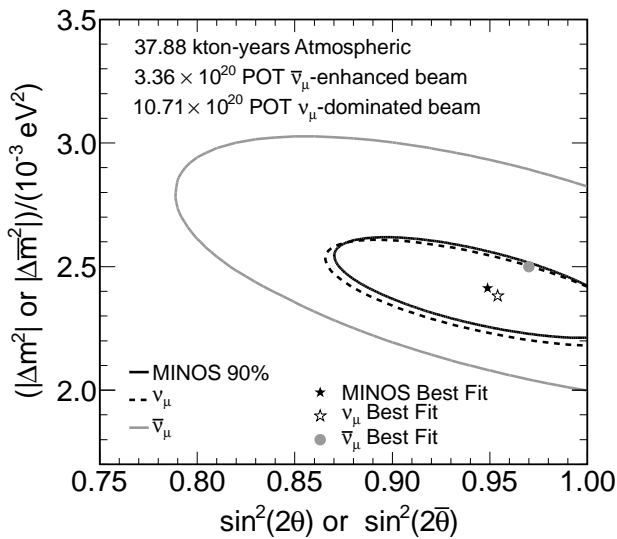}
\caption{ The 90\% confidence level allowed region of $|\dmsq|$ and
  \sinsq from the fit assuming identical neutrino and antineutrino
  oscillations (MINOS 90\%, also in Fig. \ref{fig:nu}) is compared to the allowed regions for $\numu$ and $\numubar$ from the fit in which neutrinos and antineutrinos have different oscillation parameters.}
\label{fig:nubar}
\end{figure}

The magnetized MINOS detectors enable separation of
neutrino and antineutrino interactions for both beam and atmospheric
events, allowing an independent measurement of the antineutrino
oscillation parameters.  We perform an additional fit in which we allow neutrinos
and antineutrinos to have different oscillation parameters, and find
\mbox{$|\dmbarsq|= \dmbarsqfourparall$} and \mbox{$\sinsqbar~=~\snbarfourparall$~($
> \snbarfourparalllimit$ at 90\% C.L.)}.   The difference between the
antineutrino and neutrino mass splittings is measured to be
\mbox{$\arrowvert\Delta\overline{m}^{2}\arrowvert -\arrowvert\Delta
m^{2}\arrowvert = \dmbarsqdifffourparall$}. Corresponding measurements using the beam and atmospheric samples separately yield consistent results.   The \mbox{90\% C.L.} allowed region for the antineutrino oscillation parameters is shown in Fig.~\ref{fig:nubar}, illustrating good agreement between the measured neutrino and antineutrino oscillation parameters. 

In summary, we have presented an analysis of the combined MINOS beam
and atmospheric neutrino samples, representing the complete data set
from the MINOS experiment. Assuming that neutrinos and antineutrinos
share identical oscillation parameters, we measure $\sinsq =
\sntwoparall$~($>\sntwoparalllimit$~at~90\%~C.L.) and $|\dmsq| =
\dmsqtwoparall$. Allowing independent oscillations, we measure
antineutrino parameters of $\sinsqbar =
\snbarfourparall$ ($>\snbarfourparalllimit$~at~90\%~C.L.) and
$|\dmbarsq| = \dmbarsqfourparall$. A comparison of the neutrino and
antineutrino mass splittings shows
them to be in excellent agreement. These results provide the world's most precise measurement to date of these mass splittings for both neutrinos and antineutrinos.

This work was supported by the U.S. DOE; the United Kingdom STFC; the
U.S. NSF; the State and University of Minnesota; the University of
Athens, Greece; Brazil's FAPESP, CNPq and CAPES.  We are grateful to the Minnesota Department of
Natural Resources and the personnel of the Soudan Laboratory and
Fermilab.  We thank Texas Advanced Computing Center at The University of Texas at Austin for the provision of computing resources.

\bibliographystyle{apsrev4-1}
\bibliography{paper}

\end{document}

%% file: NuMI-Atm-NuOsc-Sep12-authors.tex




\newcommand{\Berkeley}{Lawrence Berkeley National Laboratory, Berkeley, California, 94720 USA}
\newcommand{\Cambridge}{Cavendish Laboratory, University of Cambridge, Madingley Road, Cambridge CB3 0HE, United Kingdom}
\newcommand{\Cincinnati}{Department of Physics, University of Cincinnati, Cincinnati, Ohio 45221, USA}
\newcommand{\FNAL}{Fermi National Accelerator Laboratory, Batavia, Illinois 60510, USA}
\newcommand{\RAL}{Rutherford Appleton Laboratory, Science and Technologies
Facilities Council, Didcot, OX11 0QX, United Kingdom}
\newcommand{\UCL}{Department of Physics and Astronomy, University College London, Gower Street, London WC1E 6BT, United Kingdom}
\newcommand{\Caltech}{Lauritsen Laboratory, California Institute of Technology, Pasadena, California 91125, USA}
\newcommand{\Alabama}{Department of Physics and Astronomy, University of Alabama, Tuscaloosa, Alabama 35487, USA}
\newcommand{\ANL}{Argonne National Laboratory, Argonne, Illinois 60439, USA}
\newcommand{\Athens}{Department of Physics, University of Athens, GR-15771 Athens, Greece}
\newcommand{\NTUAthens}{Department of Physics, National Tech. University of Athens, GR-15780 Athens, Greece}
\newcommand{\Benedictine}{Physics Department, Benedictine University, Lisle, Illinois 60532, USA}
\newcommand{\BNL}{Brookhaven National Laboratory, Upton, New York 11973, USA}
\newcommand{\CdF}{APC -- Universit\'{e} Paris 7 Denis Diderot, 10, rue Alice Domon et L\'{e}onie Duquet, F-75205 Paris Cedex 13, France}
\newcommand{\Cleveland}{Cleveland Clinic, Cleveland, Ohio 44195, USA}
\newcommand{\Delhi}{Department of Physics \& Astrophysics, University of Delhi, Delhi 110007, India}
\newcommand{\GEHealth}{GE Healthcare, Florence South Carolina 29501, USA}
\newcommand{\Harvard}{Department of Physics, Harvard University, Cambridge, Massachusetts 02138, USA}
\newcommand{\HolyCross}{Holy Cross College, Notre Dame, Indiana 46556, USA}
\newcommand{\Houston}{Department of Physics, University of Houston, Houston, Texas 77204, USA}
\newcommand{\IIT}{Department of Physics, Illinois Institute of Technology, Chicago, Illinois 60616, USA}
\newcommand{\Iowa}{Department of Physics and Astronomy, Iowa State University, Ames, Iowa 50011 USA}
\newcommand{\Indiana}{Indiana University, Bloomington, Indiana 47405, USA}
\newcommand{\ITEP}{High Energy Experimental Physics Department, ITEP, B. Cheremushkinskaya, 25, 117218 Moscow, Russia}
\newcommand{\JMU}{Physics Department, James Madison University, Harrisonburg, Virginia 22807, USA}
\newcommand{\LASL}{Nuclear Nonproliferation Division, Threat Reduction Directorate, Los Alamos National Laboratory, Los Alamos, New Mexico 87545, USA}
\newcommand{\Lebedev}{Nuclear Physics Department, Lebedev Physical Institute, Leninsky Prospect 53, 119991 Moscow, Russia}
\newcommand{\LLL}{Lawrence Livermore National Laboratory, Livermore, California 94550, USA}
\newcommand{\LosAlamos}{Los Alamos National Laboratory, Los Alamos, New Mexico 87545, USA}
\newcommand{\Manchester}{School of Physics and Astronomy, University of Manchester, Oxford Road, Manchester M13 9PL, United Kingdom}
\newcommand{\MIT}{Lincoln Laboratory, Massachusetts Institute of Technology, Lexington, Massachusetts 02420, USA}
\newcommand{\Minnesota}{University of Minnesota, Minneapolis, Minnesota 55455, USA}
\newcommand{\Crookston}{Math, Science and Technology Department, University of Minnesota -- Crookston, Crookston, Minnesota 56716, USA}
\newcommand{\Duluth}{Department of Physics, University of Minnesota Duluth, Duluth, Minnesota 55812, USA}
\newcommand{\Ohio}{Center for Cosmology and Astro Particle Physics, Ohio State University, Columbus, Ohio 43210 USA}
\newcommand{\Otterbein}{Otterbein College, Westerville, Ohio 43081, USA}
\newcommand{\Oxford}{Subdepartment of Particle Physics, University of Oxford, Oxford OX1 3RH, United Kingdom}
\newcommand{\PennState}{Department of Physics, Pennsylvania State University, State College, Pennsylvania 16802, USA}
\newcommand{\PennU}{Department of Physics and Astronomy, University of Pennsylvania, Philadelphia, Pennsylvania 19104, USA}
\newcommand{\Pittsburgh}{Department of Physics and Astronomy, University of Pittsburgh, Pittsburgh, Pennsylvania 15260, USA}
\newcommand{\IHEP}{Institute for High Energy Physics, Protvino, Moscow Region RU-140284, Russia}
\newcommand{\Rochester}{Department of Physics and Astronomy, University of Rochester, New York 14627 USA}
\newcommand{\RoyalH}{Physics Department, Royal Holloway, University of London, Egham, Surrey, TW20 0EX, United Kingdom}
\newcommand{\Carolina}{Department of Physics and Astronomy, University of South Carolina, Columbia, South Carolina 29208, USA}
\newcommand{\SLAC}{Stanford Linear Accelerator Center, Stanford, California 94309, USA}
\newcommand{\Stanford}{Department of Physics, Stanford University, Stanford, California 94305, USA}
\newcommand{\StJohnFisher}{Physics Department, St. John Fisher College, Rochester, New York 14618 USA}
\newcommand{\Sussex}{Department of Physics and Astronomy, University of Sussex, Falmer, Brighton BN1 9QH, United Kingdom}
\newcommand{\TexasAM}{Physics Department, Texas A\&M University, College Station, Texas 77843, USA}
\newcommand{\Texas}{Department of Physics, University of Texas at Austin, 1 University Station C1600, Austin, Texas 78712, USA}
\newcommand{\TechX}{Tech-X Corporation, Boulder, Colorado 80303, USA}
\newcommand{\Tufts}{Physics Department, Tufts University, Medford, Massachusetts 02155, USA}
\newcommand{\UNICAMP}{Universidade Estadual de Campinas, IFGW-UNICAMP, CP 6165, 13083-970, Campinas, SP, Brazil}
\newcommand{\UFG}{Instituto de F\'{i}sica, Universidade Federal de Goi\'{a}s, CP 131, 74001-970, Goi\^{a}nia, GO, Brazil}
\newcommand{\USP}{Instituto de F\'{i}sica, Universidade de S\~{a}o Paulo,  CP 66318, 05315-970, S\~{a}o Paulo, SP, Brazil}
\newcommand{\Warsaw}{Department of Physics, University of Warsaw, Ho\.{z}a 69, PL-00-681 Warsaw, Poland}
\newcommand{\Washington}{Physics Department, Western Washington University, Bellingham, Washington 98225, USA}
\newcommand{\WandM}{Department of Physics, College of William \& Mary, Williamsburg, Virginia 23187, USA}
\newcommand{\Wisconsin}{Physics Department, University of Wisconsin, Madison, Wisconsin 53706, USA}
\newcommand{\deceased}{Deceased.}

\affiliation{\ANL}
\affiliation{\Athens}
\affiliation{\BNL}
\affiliation{\Caltech}
\affiliation{\Cambridge}
\affiliation{\UNICAMP}
\affiliation{\Cincinnati}
\affiliation{\FNAL}
\affiliation{\UFG}
\affiliation{\Harvard}
\affiliation{\HolyCross}
\affiliation{\Houston}
\affiliation{\IIT}
\affiliation{\Indiana}
\affiliation{\Iowa}
\affiliation{\UCL}
\affiliation{\Manchester}
\affiliation{\Minnesota}
\affiliation{\Duluth}
\affiliation{\Otterbein}
\affiliation{\Oxford}
\affiliation{\Pittsburgh}
\affiliation{\RAL}
\affiliation{\USP}
\affiliation{\Carolina}
\affiliation{\Stanford}
\affiliation{\Sussex}
\affiliation{\TexasAM}
\affiliation{\Texas}
\affiliation{\Tufts}
\affiliation{\Warsaw}
\affiliation{\WandM}

\author{P.~Adamson}
\affiliation{\FNAL}


\author{I.~Anghel}
\affiliation{\Iowa}
\affiliation{\ANL}






\author{C.~Backhouse}
\affiliation{\Oxford}




\author{G.~Barr}
\affiliation{\Oxford}









\author{M.~Bishai}
\affiliation{\BNL}

\author{A.~Blake}
\affiliation{\Cambridge}


\author{G.~J.~Bock}
\affiliation{\FNAL}


\author{D.~Bogert}
\affiliation{\FNAL}




\author{S.~V.~Cao}
\affiliation{\Texas}

\author{C.~M.~Castromonte}
\affiliation{\UFG}




% ...

\collaboration{The MINOS Collaboration}
\noaffiliation




%% file: paper.bbl
\begin{thebibliography}{26}%
\makeatletter
\providecommand \@ifxundefined [1]{%
 \@ifx{#1\undefined}
}%
\providecommand \@ifnum [1]{%
 \ifnum #1\expandafter \@firstoftwo
 \else \expandafter \@secondoftwo
 \fi
}%
\providecommand \@ifx [1]{%
 \ifx #1\expandafter \@firstoftwo
 \else \expandafter \@secondoftwo
 \fi
}%
\providecommand \natexlab [1]{#1}%
\providecommand \enquote  [1]{``#1''}%
\providecommand \bibnamefont  [1]{#1}%
\providecommand \bibfnamefont [1]{#1}%
\providecommand \citenamefont [1]{#1}%
\providecommand \href@noop [0]{\@secondoftwo}%
\providecommand \href [0]{\begingroup \@sanitize@url \@href}%
\providecommand \@href[1]{\@@startlink{#1}\@@href}%
\providecommand \@@href[1]{\endgroup#1\@@endlink}%
\providecommand \@sanitize@url [0]{\catcode `\\12\catcode `\$12\catcode
  `\&12\catcode `\#12\catcode `\^12\catcode `\_12\catcode `\%12\relax}%
\providecommand \@@startlink[1]{}%
\providecommand \@@endlink[0]{}%
\providecommand \url  [0]{\begingroup\@sanitize@url \@url }%
\providecommand \@url [1]{\endgroup\@href {#1}{\urlprefix }}%
\providecommand \urlprefix  [0]{URL }%
\providecommand \Eprint [0]{\href }%
\providecommand \doibase [0]{http://dx.doi.org/}%
\providecommand \selectlanguage [0]{\@gobble}%
\providecommand \bibinfo  [0]{\@secondoftwo}%
\providecommand \bibfield  [0]{\@secondoftwo}%
\providecommand \translation [1]{[#1]}%
\providecommand \BibitemOpen [0]{}%
\providecommand \bibitemStop [0]{}%
\providecommand \bibitemNoStop [0]{.\EOS\space}%
\providecommand \EOS [0]{\spacefactor3000\relax}%
\providecommand \BibitemShut  [1]{\csname bibitem#1\endcsname}%
\let\auto@bib@innerbib\@empty
\bibitem [{\citenamefont {Beringer}\ \emph {et~al.}(2012)\citenamefont
  {Beringer} \emph {et~al.}}]{ref:PDG2012}%
  \BibitemOpen
  \bibfield  {author} {\bibinfo {author} {\bibfnamefont {J.}~\bibnamefont
  {Beringer}} \emph {et~al.} (\bibinfo {collaboration} {Particle Data Group}),\
  }\href@noop {} {\bibfield  {journal} {\bibinfo  {journal} {Phys.\ Rev.\ D}\
  }\textbf {\bibinfo {volume} {86}},\ \bibinfo {pages} {010001} (\bibinfo
  {year} {2012})}\BibitemShut {NoStop}%
\bibitem [{ref()}]{ref:dm2note}%
  \BibitemOpen
  \href@noop {} {}\bibinfo {note} {{G. L. Fogli \etal{}, Prog.\ Part.\ Nucl.\
  Phys.\ {\bf 57}, 742 (2006); H. Nunokawa \etal{}, Phys. Rev. D {\bf 72},
  013009 (2005).}}\BibitemShut {Stop}%
\bibitem [{\citenamefont {Adamson}\ \emph
  {et~al.}(2011{\natexlab{a}})\citenamefont {Adamson} \emph
  {et~al.}}]{ref:minosCC2010}%
  \BibitemOpen
  \bibfield  {author} {\bibinfo {author} {\bibfnamefont {P.}~\bibnamefont
  {Adamson}} \emph {et~al.} (\bibinfo {collaboration} {MINOS}),\ }\href@noop {}
  {\bibfield  {journal} {\bibinfo  {journal} {Phys.\ Rev.\ Lett.}\ }\textbf
  {\bibinfo {volume} {106}},\ \bibinfo {pages} {181801} (\bibinfo {year}
  {2011}{\natexlab{a}})}\BibitemShut {NoStop}%
\bibitem [{\citenamefont {Adamson}\ \emph
  {et~al.}(2012{\natexlab{a}})\citenamefont {Adamson} \emph
  {et~al.}}]{ref:minosAtmos2012}%
  \BibitemOpen
  \bibfield  {author} {\bibinfo {author} {\bibfnamefont {P.}~\bibnamefont
  {Adamson}} \emph {et~al.} (\bibinfo {collaboration} {MINOS}),\ }\href@noop {}
  {\bibfield  {journal} {\bibinfo  {journal} {Phys.\ Rev.\ D}\ }\textbf
  {\bibinfo {volume} {86}},\ \bibinfo {pages} {052007} (\bibinfo {year}
  {2012}{\natexlab{a}})}\BibitemShut {NoStop}%
\bibitem [{\citenamefont {Adamson}\ \emph
  {et~al.}(2012{\natexlab{b}})\citenamefont {Adamson} \emph
  {et~al.}}]{ref:minosRHC_2.95}%
  \BibitemOpen
  \bibfield  {author} {\bibinfo {author} {\bibfnamefont {P.}~\bibnamefont
  {Adamson}} \emph {et~al.} (\bibinfo {collaboration} {MINOS}),\ }\href@noop {}
  {\bibfield  {journal} {\bibinfo  {journal} {Phys.\ Rev.\ Lett.}\ }\textbf
  {\bibinfo {volume} {108}},\ \bibinfo {pages} {191801} (\bibinfo {year}
  {2012}{\natexlab{b}})}\BibitemShut {NoStop}%
\bibitem [{\citenamefont {Adamson}\ \emph
  {et~al.}(2011{\natexlab{b}})\citenamefont {Adamson} \emph
  {et~al.}}]{ref:minosFHCNuBars}%
  \BibitemOpen
  \bibfield  {author} {\bibinfo {author} {\bibfnamefont {P.}~\bibnamefont
  {Adamson}} \emph {et~al.} (\bibinfo {collaboration} {MINOS}),\ }\href@noop {}
  {\bibfield  {journal} {\bibinfo  {journal} {Phys.\ Rev.\ D}\ }\textbf
  {\bibinfo {volume} {84}},\ \bibinfo {pages} {071103} (\bibinfo {year}
  {2011}{\natexlab{b}})}\BibitemShut {NoStop}%
\bibitem [{\citenamefont {{K. Anderson \etal{}}}()}]{ref:beam2}%
  \BibitemOpen
  \bibfield  {author} {\bibinfo {author} {\bibnamefont {{K. Anderson
  \etal{}}}},\ }\href@noop {} {}\bibinfo {note} {{FERMILAB-DESIGN-1998-01
  (1998).}}\BibitemShut {Stop}%
\bibitem [{\citenamefont {{For the first $3.36 \times 10^{20}$ POT of data
  taking with the $\numu$-dominated beam}}()}]{ref:HeNote}%
  \BibitemOpen
  \bibfield  {author} {\bibinfo {author} {\bibnamefont {{For the first $3.36
  \times 10^{20}$ POT of data taking with the $\numu$-dominated beam}}},\
  }\href@noop {} {}\bibinfo {note} {{the decay pipe was evacuated; it was then
  filled with helium at a pressure of 0.9 atm for structural
  reasons.}}\BibitemShut {Stop}%
\bibitem [{\citenamefont {Michael}\ \emph {et~al.}(2008)\citenamefont {Michael}
  \emph {et~al.}}]{ref:nim}%
  \BibitemOpen
  \bibfield  {author} {\bibinfo {author} {\bibfnamefont {D.~G.}\ \bibnamefont
  {Michael}} \emph {et~al.} (\bibinfo {collaboration} {MINOS}),\ }\href@noop {}
  {\bibfield  {journal} {\bibinfo  {journal} {Nucl.\ Instrum.\ Meth. A}\
  }\textbf {\bibinfo {volume} {596}},\ \bibinfo {pages} {190} (\bibinfo {year}
  {2008})}\BibitemShut {NoStop}%
\bibitem [{\citenamefont {Backhouse}(2011)}]{ref:ChrisThesis}%
  \BibitemOpen
  \bibfield  {author} {\bibinfo {author} {\bibfnamefont {C.}~\bibnamefont
  {Backhouse}},\ }\href@noop {} {\bibfield  {journal} {\bibinfo  {journal}
  {D.Phil. thesis, U. of Oxford}\ } (\bibinfo {year} {2011})}\BibitemShut
  {NoStop}%
\bibitem [{\citenamefont {Ospanov}(2008)}]{ref:RustemThesis}%
  \BibitemOpen
  \bibfield  {author} {\bibinfo {author} {\bibfnamefont {R.}~\bibnamefont
  {Ospanov}},\ }\href@noop {} {\bibfield  {journal} {\bibinfo  {journal} {Ph.D.
  thesis, U. of Texas at Austin}\ } (\bibinfo {year} {2008})}\BibitemShut
  {NoStop}%
\bibitem [{\citenamefont {Strait}(2010)}]{ref:MattStraitThesis}%
  \BibitemOpen
  \bibfield  {author} {\bibinfo {author} {\bibfnamefont {M.~L.}\ \bibnamefont
  {Strait}},\ }\href@noop {} {\bibfield  {journal} {\bibinfo  {journal} {Ph.D.
  thesis, U. of Minnesota}\ } (\bibinfo {year} {2010})}\BibitemShut {NoStop}%
\bibitem [{\citenamefont {McGowan}(2007)}]{ref:AaronThesis}%
  \BibitemOpen
  \bibfield  {author} {\bibinfo {author} {\bibfnamefont {A.}~\bibnamefont
  {McGowan}},\ }\href@noop {} {Ph.D. thesis},\ \bibinfo  {school} {U. of
  Minnesota} (\bibinfo {year} {2007})\BibitemShut {NoStop}%
\bibitem [{\citenamefont {Adamson}\ \emph {et~al.}(2008)\citenamefont {Adamson}
  \emph {et~al.}}]{ref:minos2008}%
  \BibitemOpen
  \bibfield  {author} {\bibinfo {author} {\bibfnamefont {P.}~\bibnamefont
  {Adamson}} \emph {et~al.} (\bibinfo {collaboration} {MINOS}),\ }\href@noop {}
  {\bibfield  {journal} {\bibinfo  {journal} {Phys.\ Rev.\ D}\ }\textbf
  {\bibinfo {volume} {77}},\ \bibinfo {pages} {072002} (\bibinfo {year}
  {2008})}\BibitemShut {NoStop}%
\bibitem [{\citenamefont {Barr}\ \emph {et~al.}(2004)\citenamefont {Barr} \emph
  {et~al.}}]{ref:Bartol}%
  \BibitemOpen
  \bibfield  {author} {\bibinfo {author} {\bibfnamefont {G.~D.}\ \bibnamefont
  {Barr}} \emph {et~al.},\ }\href@noop {} {\bibfield  {journal} {\bibinfo
  {journal} {Phys.\ Rev.\ D}\ }\textbf {\bibinfo {volume} {70}},\ \bibinfo
  {pages} {023006} (\bibinfo {year} {2004})}\BibitemShut {NoStop}%
\bibitem [{\citenamefont {Battistoni}\ \emph {et~al.}(2007)\citenamefont
  {Battistoni} \emph {et~al.}}]{ref:FlukaAIP}%
  \BibitemOpen
  \bibfield  {author} {\bibinfo {author} {\bibfnamefont {G.}~\bibnamefont
  {Battistoni}} \emph {et~al.},\ }\href@noop {} {\bibfield  {journal} {\bibinfo
   {journal} {AIP Conf. Proc.}\ }\textbf {\bibinfo {volume} {896}},\ \bibinfo
  {pages} {31} (\bibinfo {year} {2007})}\BibitemShut {NoStop}%
\bibitem [{\citenamefont {Amako}\ \emph {et~al.}(2006)\citenamefont {Amako}
  \emph {et~al.}}]{ref:GEANT4IEEE}%
  \BibitemOpen
  \bibfield  {author} {\bibinfo {author} {\bibfnamefont {K.}~\bibnamefont
  {Amako}} \emph {et~al.} (\bibinfo {collaboration} {GEANT4}),\ }\href@noop {}
  {\bibfield  {journal} {\bibinfo  {journal} {IEEE Trans. on Nucl. Sci.}\
  }\textbf {\bibinfo {volume} {53}},\ \bibinfo {pages} {270} (\bibinfo {year}
  {2006})}\BibitemShut {NoStop}%
\bibitem [{\citenamefont {Gallagher}(2002)}]{ref:Neugen}%
  \BibitemOpen
  \bibfield  {author} {\bibinfo {author} {\bibfnamefont {H.}~\bibnamefont
  {Gallagher}},\ }\href@noop {} {\bibfield  {journal} {\bibinfo  {journal}
  {Nucl.\ Phys.\ Proc.\ Suppl.}\ }\textbf {\bibinfo {volume} {112}},\ \bibinfo
  {pages} {188} (\bibinfo {year} {2002})}\BibitemShut {NoStop}%
\bibitem [{\citenamefont {Casper}(2002)}]{ref:Nuance}%
  \BibitemOpen
  \bibfield  {author} {\bibinfo {author} {\bibfnamefont {D.}~\bibnamefont
  {Casper}},\ }\href@noop {} {\bibfield  {journal} {\bibinfo  {journal} {Nucl.\
  Phys.\ Proc.\ Suppl.}\ }\textbf {\bibinfo {volume} {112}},\ \bibinfo {pages}
  {161} (\bibinfo {year} {2002})}\BibitemShut {NoStop}%
\bibitem [{\citenamefont {Zeitnitz}\ and\ \citenamefont
  {Gabriel}(1994)}]{ref:GCALOR}%
  \BibitemOpen
  \bibfield  {author} {\bibinfo {author} {\bibfnamefont {C.}~\bibnamefont
  {Zeitnitz}}\ and\ \bibinfo {author} {\bibfnamefont {T.~A.}\ \bibnamefont
  {Gabriel}},\ }\href@noop {} {\bibfield  {journal} {\bibinfo  {journal}
  {Nucl.\ Instrum.\ Meth.\ A}\ }\textbf {\bibinfo {volume} {349}},\ \bibinfo
  {pages} {106} (\bibinfo {year} {1994})}\BibitemShut {NoStop}%
\bibitem [{\citenamefont {{Application Software Group}}(1994)}]{ref:GEANT3}%
  \BibitemOpen
  \bibfield  {author} {\bibinfo {author} {\bibnamefont {{Application Software
  Group}}},\ }\href@noop {} {}\bibinfo {type} {CERN Program Library Long
  Writeup}\ \bibinfo {number} {W5013}\ (\bibinfo  {institution} {CERN},\
  \bibinfo {year} {1994})\BibitemShut {NoStop}%
\bibitem [{\citenamefont {Coleman}(2011)}]{ref:StephenThesis}%
  \BibitemOpen
  \bibfield  {author} {\bibinfo {author} {\bibfnamefont {S.~J.}\ \bibnamefont
  {Coleman}},\ }\href@noop {} {Ph.D. thesis},\ \bibinfo  {school} {Coll. of
  William \& Mary} (\bibinfo {year} {2011})\BibitemShut {NoStop}%
\bibitem [{\citenamefont {Mitchell}(2011)}]{ref:JessThesis}%
  \BibitemOpen
  \bibfield  {author} {\bibinfo {author} {\bibfnamefont {J.}~\bibnamefont
  {Mitchell}},\ }\href@noop {} {Ph.D. thesis},\ \bibinfo  {school} {U. of
  Cambridge} (\bibinfo {year} {2011})\BibitemShut {NoStop}%
\bibitem [{\citenamefont {Blake}\ \emph {et~al.}(2013)\citenamefont {Blake}
  \emph {et~al.}}]{ref:minosAtmos2012NIM}%
  \BibitemOpen
  \bibfield  {author} {\bibinfo {author} {\bibfnamefont {A.}~\bibnamefont
  {Blake}} \emph {et~al.},\ }\href@noop {} {\bibfield  {journal} {\bibinfo
  {journal} {Nucl. Instrum. Meth. A}\ }\textbf {\bibinfo {volume} {707}},\
  \bibinfo {pages} {127} (\bibinfo {year} {2013})}\BibitemShut {NoStop}%
\bibitem [{\citenamefont {Itow}()}]{ref:SKNeutrino2012}%
  \BibitemOpen
  \bibfield  {author} {\bibinfo {author} {\bibfnamefont {Y.}~\bibnamefont
  {Itow}},\ }in\ \href@noop {} {\emph {\bibinfo {booktitle} {Proceedings of the
  25th International Conference on Neutrino Physics and Astrophysics (Neutrino
  2012), Kyoto, Japan, June 2012}}},\ \bibinfo {note} {to be
  published}\BibitemShut {NoStop}%
\bibitem [{\citenamefont {Abe}\ \emph {et~al.}(2012)\citenamefont {Abe} \emph
  {et~al.}}]{ref:T2Knucontour}%
  \BibitemOpen
  \bibfield  {author} {\bibinfo {author} {\bibfnamefont {K.}~\bibnamefont
  {Abe}} \emph {et~al.} (\bibinfo {collaboration} {T2K}),\ }\href@noop {}
  {\bibfield  {journal} {\bibinfo  {journal} {Phys.\ Rev.\ D}\ }\textbf
  {\bibinfo {volume} {85}},\ \bibinfo {pages} {031103} (\bibinfo {year}
  {2012})}\BibitemShut {NoStop}%
\end{thebibliography}%
